arXiv:2002.10227

# The nonextensive statistical ensembles with dual thermodynamic interpretations

Yahui Zheng[1], Jiulin Du[2], Linxia Liu[3], Huijun Kong[1]

1: *Department of Electronics and Communication Engineering, Henan Institute of Technology, Xinxiang City 453003, China.*

2: *Department of Physics, School of Science, Tianjin University, Tianjin 300072, China.*

3 *Teaching Research and Assessment Center, Henan Institute of Technology, Xinxiang 453003, China.***Abstract:** The nonextensive statistical ensembles are revisited for the complex systems with long-range interactions and long-range correlations. An approximation, the value of nonextensive parameter (1-$q$) is assumed to be very tiny, is adopted for the limit of large particle number for most normal systems. In this case, Tsallis entropy can be expanded as a function of energy and particle number fluctuation, and thus the power-law forms of the generalized Gibbs distribution and grand canonical distribution can be derived. These new distribution functions can be applied to derive the free energy and grand thermodynamic potential in nonextensive thermodynamics. In order to establish appropriate nonextensive thermodynamic formalism, the dual thermodynamic interpretations are necessary for thermodynamic relations and thermodynamic quantities. By using a new technique of parameter transformation, the single-particle distribution can be deduced from the power-law Gibbs distribution. This technique produces a link between the statistical ensemble and the quasi-independent system with two kinds of nonextensive parameter having quite different physical explanations. Furthermore, the technique is used to construct nonextensive quantum statistics and effectively to avoid the factorization difficulty in the power-law grand canonical distribution.

**Keywords:** Nonextensive statistical ensemble, Nonextensive thermodynamics, Power-law distribution, Complex system## 1 Introduction

In recent years, nonextensive statistical mechanics (NSM) has been developed to deal with the complex systems, such as the systems with long-range interactions and long-range correlations, with the fractal phase space/time space, and with long-range memory effect etc. This new statistical theory is based on the $q$-entropy proposed by Tsallis in 1988 [1]. By using the maximum entropy principle for the $q$-entropy, the power-law $q$-distribution functions are found, which are equal to Boltzmann-Gibbs distribution only when the $q$-parameter is equal to unity. NSM has been widely applied to many interesting research fields in physics, astronomy, chemistry, life science and technology, the representative examples such as the self-gravitating systems [2-8], the astrophysical and space plasmas [9-12], the anomalous diffusion systems [13-15], biological and chemical reaction systems [16-20] and so on.

With the development of NSM, there have been at least four kinds of formulism so far. They include the original method [1], un-normalized method [21], normalized method [22], and the optimal Lagrange multiplier (OLM) method [23]. To avoid the self-referential problem, recently one generalized version of OLM method was proposed by an assumption that the Tsallis factor is independent of probability distribution of each microscopic configuration [24]. In the fundamental nonextensive thermodynamics [25-28], there have been some obstacles in establishment of complete nonextensive thermodynamic formalism. For example, temperature can not be naturally defined in the zeroth law of thermodynamics, but it can be defined in a modified manner in fundamental thermodynamic equations of the first law. And free energy is also modified by adding the Tsalis factor [29] so that unfamiliar thermodynamic relations are produced in nonextensive thermodynamics. This is not consistent with our physical intuition, where the thermodynamic laws come from the phenomenological observations and experiments, and they are appropriate for any systems, no matter whatever statistical mechanics the systems are based on. Recently, a temperature duality assumption was given to solve the problem of temperature definition [41,46], where two parallel thermodynamic equations were put forward in the first law of thermodynamics and therefore the dual physical interpretations were given for the thermodynamic quantities, such as temperature, pressure, internal energy, free energy and so on. Thus the nonextensive thermodynamic formalism was proposed having two parallel Legendre transformation structures.

In nonextensive statistical ensemble theory on the complex systems, some fundamental works had been performed, for example, the generalized power-law Gibbs distributions based on the dynamical thermostatting approach [30], counting rule [31], the fractal dimension of phase space [32], and the order parameter in the generalized Lorentzian [33]. Recently, the nonextensive statistical ensemble has been restudied based on the traditional method employing the assumption of equiprobability [34,35], where the interactions between a system and its reservoir are short-range and the entropy is extensive. Therefore, more complex mechanism for the complex systems with long-range interactions has not been studied. And the thermodynamic treatment is still taken root in single definition for temperature, and therefore the applicable nonextensive thermodynamic formalism can hardly be established.

In this work, based on the equiprobability assumption and the dual physical interpretations of thermodynamic equations and thermodynamic quantities, we study the nonextensive statistical ensemble approach for the complex systems with long-range interactions.

The paper is organized as follows. In section 2, the micro-canonical $q$-distribution and basic nonextensive thermodynamic equations are discussed. In section 3, the Gibbs $q$-distribution in the canonical ensemble and the link of the dual internal energies in NSM are studied. In section 4, the grand canonical $q$-distribution and the link of the dual particle numbers in NSM are analyzed. In section 5, the single-particle $q$-distribution function based on a parameter transformation is proposed in the canonical ensemble. In section 6, the nonextensive quantum statistics on the grand canonical ensemble is revisited. In section 7, conclusions and discussions are given.

## 2 The nonextensive canonical ensembles with equiprobability

In complex systems, generally speaking, the ergodic assumption in the phase space is violated due to the long-range interactions, therefore some phase points cannot be felt by the fundamental dynamic process and some microstates cannot be arrived in the evolution series. However, we can still assume that in the confined phase space, all the available microstates or the quantum states have an equal opportunity to appear in the dynamic evolution processes.

In this situation, the $q$-entropy of an isolated complex system is considered as the form of $q$-logarithm,

$$S_q = k \ln_q \Omega \equiv k \frac{\Omega^{1-q} - 1}{1 - q}, \tag{1}$$

where $\Omega$ is the number of quantum states and the $q$-logarithm is $\ln_q x = (x^{1-q}-1)/(1-q)$. Usually, we introduce Tsallis factor as

$$c_q \equiv \frac{S_q}{k}(1-q)+1. \tag{2}$$

In the case of equiprobability, it is written as $c_q = \Omega^{1-q}$.

In order to study the canonical ensemble in NSM, let us consider a composite isolated system consisting of the observed system and its reservoir. Inside the whole composite system, there exists long-range interactions, and between the observed system and its reservoir there are also the interactions. As the beginning, we first consider the composition rule of the energy level at different quantum states of the whole composite system. Due to the long-range interactions, there must be long-range correlations between different quantum states, which can be described by

$$E_{ij(1,2)} = E_{i1} + E_{j2} - (1-q)\frac{\beta}{c_q}(E_{i1} - U_{q1}^{(3)})(E_{j2} - U_{q2}^{(3)}), \tag{3}$$

where the subscripts 1 and 2 represent different subsystems, and the "*i*" and "*j*" represent different quantum states, respectively. In the composition rule (3), *β* is the Lagrange multiplier and *U* is internal energy or an averaged value of the energy of some given system, which will be defined later. The cross term in Eq. (3) explicitly shows the long-range correlations between different quantum states. It is obvious that there exists energy fluctuation for different quantum states in this cross term, which is generally defined as

$$\Delta E = E - U_q^{(3)}, \tag{4}$$

where *E* is the energy level on some quantum state. Here the superscript "(3)" denotes the third choice of average definition [22] in NSM. We will show that (1-*q*) is inversely proportional to the particle number, from which it is easy to find that the value of (1-*q*) is very tiny when particle number is large enough. Furthermore, the square of energy fluctuation in Eq. (3) is ordinarily proportional to particle number. Therefore, in the limit of large particle number, the cross term in Eq. (3) is ignorable.

Then the composite rule of the quantum states becomes additive, namely,

$$E_{ij(1,2)} = E_{i1} + E_{j2}. \tag{5}$$

The above rule holds in the case of large particle number, especially in the systems having no phase transitions. According to this rule, the total energy $E_0$ of the composite system can be written as sum of the system energy $E_s$ at quantum state *s* and the reservoir energy $E_r$,

$$E_0 = E_s + E_r. \tag{6}$$

When the observed system is at quantum state (or microstate) *s*, the possible quantum state (or microstate) number for the composite system is $\Omega_r(E_0-E_s)$, which, according to the equal-probability principle, is proportional to the probability distribution $\rho_s$ of the system at microstate *s*. So we have that

$$\rho_s = \frac{\Omega_r(E_0 - E_s)}{\Omega_t(E_0)}, \tag{7}$$

where $\Omega_t(E_0)$ is the total number of possible quantum states of the isolated composite system.

We introduce two quantities $U_q^{(3)}$ and $U_{qr}^{(3)}$ to represent the internal energies of the system and reservoir, respectively, which are equal to the values of the most probable energies at the *q*-equilibrium state. Then according to Eq.(4), the energy fluctuations of the system and reservoir can be given, which are both much less than internal energies of the system and the reservoir, respectively. Therefore, around the most probable internal energy of reservoir, the quantity $\ln_q\Omega_r$ can be expanded as a function of the energy fluctuation of the observed system. Only taking the first two terms, we have that

$$\ln_q \Omega_r(E_0 - E_s) = \ln_q \Omega_r(E_0 - U_q^{(3)}) - \left(\frac{\partial \ln_q \Omega_r}{\partial E_r}\right)_{E_r = E_0 - U_q^{(3)}} (E_s - U_q^{(3)})$$
$$= \ln_q \Omega_r(U_{qr}^{(3)}) - \beta_r(E_s - U_q^{(3)}), \quad (8)$$

where the Lagrange multiplier is introduced through

$$\beta = \frac{\partial \ln_q \Omega}{\partial U_q^{(3)}}. \quad (9)$$

It is interesting that the expansion in Eq. (8) does not depend on the assumption of large reservoir. The thermal balance condition for the canonical ensemble can be expressed as

$$\frac{\beta}{c_q} = \frac{\beta_r}{c_{qr}}. \quad (10)$$

So the probability (7) can be written (see Appendix) by

$$\rho_s = \frac{1}{\bar{Z}_q^{(3)}}[1 - (1-q)\frac{\beta}{c_q}(E_s - U_q^{(3)})]^{\frac{1}{1-q}}, \quad (11)$$

where the partition function is defined by

$$\bar{Z}_q^{(3)} = \frac{\Omega_t(E_0)}{\Omega_r(U_{qr}^{(3)})} = \sum_s [1 - (1-q)\frac{\beta}{c_q}(E_s - U_q^{(3)})]^{\frac{1}{1-q}}. \quad (12)$$

This probability distribution function (11) is the generalized Gibbs power-law $q$-distribution, which is exactly identical to that obtained along the procedure of entropy maximization [16]. According to Eq. (11), Tsallis entropy of the observed system is expressed as

$$S_q = k \frac{\sum \rho_s^q - 1}{1 - q}, \quad (13)$$

from which Tsallis factor of the system is given by

$$c_q = \sum_s \rho_s^q. \quad (14)$$

It can be proved that for the partition function (12), we have that

$$\sum_s \rho_s^q = [\bar{Z}_q^{(3)}]^{1-q}. \quad (15)$$

Furthermore, one can find that the probability independence is guaranteed in the composite rule (3). We introduce the definition of the internal energy formally and we employ the third choice of average definition [22] in NSM, namely,

$$U_q^{(3)} \equiv \frac{\sum E_s \rho_s^q}{\sum \rho_s^q}. \quad (16)$$

It can be seen that with both the two rules (3) and (4), the additivity of internal energy can be reserved, namely,

$$U_{q1,2}^{(3)} = U_{q1}^{(3)} + U_{q2}^{(3)}, \quad (17)$$

where the subscript 1 and 2 represent the system and the reservoir respectively. According to Eqs.(11), (13) and (16), the fundamental thermodynamic equation can be given [36] by

$$dU_q^{(3)} = TdS_q - PdV, \quad (18)$$

where the Lagrange temperature is introduced by

$$\beta = \frac{1}{kT}, \quad (19)$$

where $P$ is the system pressure and $V$ is its volume. From Eq.(18) we can obtain the following thermodynamic relations,

$$T = \left(\frac{\partial U_q^{(3)}}{\partial S_q}\right)_V, \quad -P = \left(\frac{\partial U_q^{(3)}}{\partial V}\right)_{S_q}. \quad (20)$$

They can also be regarded as the definition of the temperature and the pressure in nonextensive thermodynamics.

Due to the self-referential property, it is difficult to apply the $q$-distribution function (11) to realistic complex systems. So we need another form. For this purpose, we employ that

$$\rho_s = \frac{1}{\bar{Z}_q^{(3)}}[1+(1-q)\frac{\beta}{c_q}U_q^{(3)}]^{\frac{1}{1-q}}[1-(1-q)\frac{\beta}{c_q+(1-q)\beta U_q^{(3)}}E_s]^{\frac{1}{1-q}}. \quad (21)$$

In light of tiny value of $(1-q)$, we have the approximation:

$$\rho_s \simeq \frac{1}{\bar{Z}_q^{(3)}[1+(1-q)\frac{\beta}{c_q}U_q^{(3)}]^{\frac{-1}{1-q}}}[1-(1-q)\frac{\beta}{c_q}E_s]^{\frac{1}{1-q}}. \quad (22)$$

Now we define a new partition function,

$$Z_q^{(3)} \equiv \sum_s [1-(1-q)\frac{\beta}{c_q}E_s]^{\frac{1}{1-q}}, \quad (23)$$

which [24] should satisfy

$$Z_q^{(3)} = \bar{Z}_q^{(3)}[1+(1-q)\frac{\beta}{c_q}U_q^{(3)}]^{\frac{-1}{1-q}}. \quad (24)$$

Then the $q$-distribution function (11) is changed as

$$\rho_s = \frac{1}{Z_q^{(3)}}[1-(1-q)\frac{\beta}{c_q}E_s]^{\frac{1}{1-q}}. \quad (25)$$

The above distribution can also be calculated on basis of the procedure of entropy maximization by assuming that the partial derivative of Tsallis factor to the probability distribution function of some configuration is zero [24].

Furthermore, from Eq. (23) and in view of Eq. (15), one can directly obtain that

$$c_q = [Z_q^{(3)}]^{1-q} + (1-q)\frac{\beta}{c_q}[Z_q^{(3)}]^{1-q}U_q^{(3)} \approx [Z_q^{(3)}]^{1-q} + (1-q)\beta U_q^{(3)}. \quad (26)$$

The above result could be exactly calculated on basis of the modified distribution (25) together with the definition (16). Then the free energy in the representation (16) is defined as

$$F_q^{(3)} \equiv U_q^{(3)} - TS_q = -kT \ln_q Z_q^{(3)}, \quad (27)$$

On the basis of Eq. (25), we have the following statistical expression of internal energy [24],

$$U_q^{(3)} = -\frac{\partial}{\partial \beta}\ln_q Z_q^{(3)}. \quad (28)$$

In next section, we shall study the grand canonical ensemble in the nonextensive statistics based on the assumption of equiprobability.

## 3 The nonextensive grand canonical ensembles with equiprobability

For the grand canonical ensemble, there are energy and particles exchange between the system and its reservoir. Similarly, we also consider an isolated composite system, consisting of the observed system and the reservoir. For the composite system with long-range interactions, the composition rules of energy level and particle numbers at different microstates are, respectively,

$$\alpha N_{ij(1,2)} + \beta E_{ij(1,2)} = \alpha N_{i1} + \beta E_{i1} + \alpha N_{j2} + \beta E_{j2}$$
$$- \frac{(1-q)}{c_q}[\alpha(N_{i1} - \bar{N}_1^{(3)}) + \beta(E_{i1} - U_{q1}^{(3)})][\alpha(N_{j2} - \bar{N}_2^{(3)}) + \beta(E_{j2} - U_{q2}^{(3)})], \quad (29)$$

where $\bar{N}^{(3)}$ is the average particle number based on the third choice of average definition [22], the subscript 1 and 2 denote different subsystems, and the subscripts "$i$" and "$j$" represent different microstates. The quantities $\alpha$ and $\beta$ are the Lagrange multipliers which will be defined later.

Similarly, we can confirm that for an open system, the parameter (1-$q$) is inversely proportional to the averaged particle number. On the other hand, either the square of energy fluctuation or particle number fluctuation, or the product of energy fluctuation and particle number fluctuation in Eq. (29), is proportional to the averaged particle number. This means that in the limit of large particle number, the cross term of Eq. (29) can be ignored. Therefore, we now adopt the following additive rules,

$$\alpha N_{ij(1,2)} + \beta E_{ij(1,2)} = \alpha N_{i1} + \beta E_{i1} + \alpha N_{j2} + \beta E_{j2}. \quad (30)$$

Now that the Lagrange multipliers are indefinite, we generally have

$$E_{ij(1,2)} = E_{i1} + E_{j2}, \quad N_{ij(1,2)} = N_{i1} + N_{j2}. \quad (31)$$

The above rules hold for the complex systems without any phase transition. According to Eq. (31), the total energy $E_0$ of the composite system can be written as the sum of the system energy $E_s$ at the microstate $s$ and the reservoir energy $E_r$. And meanwhile, the total particle number of the composite system $N_0$ is sum of the particle number of the observed system $N_s$ and the reservoir $N_r$. That is,

$$E_0 = E_s + E_r, \quad N_0 = N_s + N_r. \quad (32)$$

According to the equal probability principle, the probability distribution of the system with energy $E_s$ and particle number $N_s$ is proportional to state number of the reservoir with energy $E_r$ and particle number $N_r$, i.e.,

$$\rho_{Ns} = \frac{\Omega_r(E_0 - E_s, N_0 - N)}{\Omega_t(E_0, N_0)}. \quad (33)$$

Accordingly, we can define the particle number fluctuation and the energy fluctuation of the observed system by

$$\Delta N = N_s - \bar{N}^{(3)}, \quad \Delta E = E_s - U_q^{(3)}. \quad (34)$$

At the $q$-equilibrium state, the energy fluctuation and particle fluctuation are both very small relatively to the most probable energies and particle numbers of the observed system and reservoir. Therefore, the quantity $\ln_q \Omega_r$ can be regarded as a function of the particle number fluctuation and the energy fluctuation of the observed system and can be expanded. Only taking the first three

terms, we have that

$$\ln_q \Omega_r(N_0 - N_s, E_0 - E_s)$$
$$= \ln_q \Omega_r(N_0 - \bar{N}^{(3)}, E_0 - U_q^{(3)}) - \left(\frac{\partial \ln_q \Omega_r}{\partial N_r}\right)_{N_r = \bar{N}_r^{(3)}} (N_s - \bar{N}^{(3)}) - \left(\frac{\partial \ln_q \Omega_r}{\partial E_r}\right)_{E_r = U_{qr}^{(3)}} (E_s - U_q^{(3)})$$
$$= \ln_q \Omega_r(\bar{N}_r^{(3)}, U_{qr}^{(3)}) - \alpha_r N_s - \beta_r E_s, \tag{35}$$

where the Lagrange multipliers are defined as

$$\alpha = \frac{\partial \ln_q \Omega}{\partial \bar{N}^{(3)}}, \quad \beta = \frac{\partial \ln_q \Omega}{\partial U_q^{(3)}}. \tag{36}$$

The balance condition for the grand canonical ensemble can be expressed as

$$\frac{\beta}{c_q} = \frac{\beta_r}{c_{qr}}, \quad \frac{\alpha}{c_q} = \frac{\alpha_r}{c_{qr}}. \tag{37}$$

Then the grand ensemble probability $q$-distribution can be derived by

$$\rho_{Ns} = \frac{1}{\bar{\Xi}_q^{(3)}} [1 - (1-q)[\frac{\alpha}{c_q}(N_s - \bar{N}^{(3)}) + \frac{\beta}{c_q}(E_s - U_q^{(3)})]]^{\frac{1}{1-q}}$$
$$\equiv \frac{1}{\bar{\Xi}_q^{(3)}} e_q^{-[\frac{\alpha}{c_q}(N_s - \bar{N}^{(3)}) + \frac{\beta}{c_q}(E_s - U_q^{(3)})]}, \tag{38}$$

where the grand partition function is defined by

$$\bar{\Xi}_q^{(3)} = \frac{\Omega_t(E_0, N_0)}{\Omega_r(\bar{N}_r^{(3)}, U_{qr}^{(3)})} = \sum_{N=0}^{\infty} \sum_s e_q^{-[\frac{\alpha}{c_q}(N_s - \bar{N}^{(3)}) + \frac{\beta}{c_q}(E_s - U_q^{(3)})]}. \tag{39}$$

On the basis of the distribution (38), Tsallis entropy is expressed as

$$S_q = k \frac{\sum_{N=0}^{\infty} \sum_s \rho_{Ns}^q - 1}{1-q}, \tag{40}$$

and the Tsallis factor is given by

$$c_q = \sum_{N=0}^{\infty} \sum_s \rho_{Ns}^q = [\bar{\Xi}_q^{(3)}]^{1-q}. \tag{41}$$

It is obvious that the grand canonical $q$-distribution (38) satisfies the probability independence based on the composite rule (29). Now definitions of the internal energy and the average particle number can be given by

$$U_q^{(3)} = \frac{\sum_{N=0}^{\infty} \sum_s \rho_{Ns}^q E_s}{\sum_{N=0}^{\infty} \sum_s \rho_{Ns}^q}, \quad \bar{N}^{(3)} = \frac{\sum_{N=0}^{\infty} \sum_s \rho_{Ns}^q N_s}{\sum_{N=0}^{\infty} \sum_s \rho_{Ns}^q}. \tag{42}$$

Then from (29) or (31), we have the following additive relations,

$$U_{q1,2}^{(3)} = U_{q1}^{(3)} + U_{q2}^{(3)}, \quad \bar{N}_{1,2}^{(3)} = \bar{N}_1^{(3)} + \bar{N}_2^{(3)}, \tag{43}$$

where the subscripts 1 and 2 represent the system and the reservoir, respectively.

Now we deduce the fundamental thermodynamic equation in the grand canonical ensemble in the representation of Eq. (42). Firstly, let us introduce the following Lagrange relations,

$$\beta = \frac{1}{kT}, \quad \alpha = -\mu\beta = -\frac{\mu}{kT}, \tag{44}$$

where $\mu$ is the chemical potential of the system. According to Eq. (42), one has [36] that

$$dU_q^{(3)} = \frac{q\sum_{N=0}^{\infty}\sum_s \rho_{Ns}^{q-1}(E_s - U_q^{(3)})d\rho_{Ns}}{c_q} + \frac{\sum_{N=0}^{\infty}\sum_s \rho_{Ns}^q dE_s}{c_q}, \tag{45}$$

$$d\bar{N}^{(3)} = \frac{q\sum_{N=0}^{\infty}\sum_s \rho_{Ns}^{q-1}(N_s - \bar{N}^{(3)})d\rho_{Ns}}{c_q} + \frac{\sum_{N=0}^{\infty}\sum_s \rho_{Ns}^q dN_s}{c_q}. \tag{46}$$

Combining Eqs. (45) and (46) with the chemical potential, one gets

$$dU_q^{(3)} - \mu d\bar{N}^{(3)} = q\sum_{N=0}^{\infty}\sum_s \rho_{Ns}^{q-1} c_q^{-1}\left[(E_s - U_q^{(3)}) - \mu(N_s - \bar{N}^{(3)})\right]d\rho_{Ns}$$
$$+ c_q^{-1}\sum_{N=0}^{\infty}\sum_s \rho_{Ns}^q d(E_s - \mu N_s). \tag{47}$$

On the right-hand side of the above equation, the first term is heat absorbed by the system from the reservoir, namely,

$$dQ^{(3)} = q\sum_{N=0}^{\infty}\sum_s \rho_{Ns}^{q-1} c_q^{-1}\left[(E_s - U_q^{(3)}) - \mu(N_s - \bar{N}^{(3)})\right]d\rho_{Ns}. \tag{48}$$

Based on the grand canonical $q$-distribution (38), one can derive that

$$q\sum_{N=0}^{\infty}\sum_s \rho_{Ns}^{q-1} c_q^{-1}\left[(E_s - U_q^{(3)}) - \mu(N_s - \bar{N}^{(3)})\right]d\rho_{Ns}$$
$$= \frac{qkTc_q}{1-q}\sum_{N=0}^{\infty}\sum_s [1-(1-q)\frac{(E_s - U_q^{(3)}) - \mu(N_s - \bar{N}^{(3)})}{c_q kT}]^{-1}d\rho_{Ns}. \tag{49}$$

On the other hand, in light of Eqs.(38) and (40), we have

$$dS_q = \frac{qkc_q}{1-q}\sum_{N=0}^{\infty}\sum_s [1-(1-q)\frac{(E_s - U_q^{(3)}) - \mu(N_s - \bar{N}^{(3)})}{kc_q T}]^{-1}d\rho_{Ns}. \tag{50}$$

Therefore, there is the relation,

$$dQ^{(3)} = TdS_q. \tag{51}$$

Furthermore, the second term on the right-hand side of Eq. (47) is the work done by the reservoir on the system in the grand canonical ensemble, namely,

$$dW^{(3)} = \frac{\sum_{N=0}^{\infty}\sum_s \rho_{Ns}^q d(E_s - \mu N_s)}{c_q} = -PdV. \tag{52}$$

Substituting Eqs.(51) and (52) into Eq. (47), we immediately obtain

$$dU_q^{(3)} = TdS_q - PdV + \mu d\bar{N}^{(3)}. \tag{53}$$

From Eq. (53), we find the following relations:

$$T = \left(\frac{\partial U_q^{(3)}}{\partial S_q}\right)_{V,\bar{N}^{(3)}}, \quad -P = \left(\frac{\partial U_q^{(3)}}{\partial V}\right)_{S_q,\bar{N}^{(3)}}, \quad \mu = \left(\frac{\partial U_q^{(3)}}{\partial \bar{N}^{(3)}}\right)_{S_q,V}. \tag{54}$$

Again, due to the self-referential property, it is difficult for the grand canonical $q$-distribution (38) to be applied to realistic complex systems. Now we make the modification,

$$\rho_{Ns} = \frac{1}{\bar{\Xi}_q^{(3)}}[1+(1-q)\frac{\alpha\bar{N}^{(3)}+\beta U_q^{(3)}}{c_q}]^{\frac{1}{1-q}}[1-(1-q)\frac{\alpha N_s + \beta E_s}{c_q + (1-q)(\alpha\bar{N}^{(3)}+\beta U_q^{(3)})}]^{\frac{1}{1-q}}$$

$$\approx \frac{1}{\bar{\Xi}_q^{(3)}[1+(1-q)\frac{U_q^{(3)}-\mu\bar{N}^{(3)}}{kc_qT}]^{\frac{-1}{1-q}}}[1-(1-q)\frac{E_s - \mu N_s}{kc_qT}]^{\frac{1}{1-q}}, \tag{55}$$

where the Lagrange relations (44) have been considered. We define new grand partition function:

$$\Xi_q^{(3)} \equiv \sum_{N=0}^{\infty}\sum_s [1-(1-q)\frac{E_s - \mu N_s}{kc_qT}]^{\frac{1}{1-q}}. \tag{56}$$

Then the $q$-distribution (38) becomes

$$\rho_{Ns} = \frac{1}{\Xi_q^{(3)}}[1-(1-q)\frac{E_s - \mu N_s}{kc_qT}]^{\frac{1}{1-q}}. \tag{57}$$

This $q$-distribution function can also be deduced in the maximization entropy procedure by use of the assumption that Tsallis factor is irrelevant to the $q$-distribution function itself [24].

The new grand partition function in Eq. (57) should satisfy

$$\Xi_q^{(3)} = \bar{\Xi}_q^{(3)}[1+(1-q)\frac{U_q^{(3)}-\mu\bar{N}^{(3)}}{kc_qT}]^{\frac{-1}{1-q}}. \tag{58}$$

Based on Eq. (41), we have the following approximate relation,

$$c_q \simeq [\Xi_q^{(3)}]^{1-q} + (1-q)\frac{U_q^{(3)}-\mu\bar{N}^{(3)}}{kT}. \tag{59}$$

The above relation can also exactly be calculated through Eqs. (42), (56) and (57). Then the free energy in the grand canonical ensemble in the representation of Eq. (42) is

$$F_q^{(3)} \equiv U_q^{(3)} - TS_q = -kT\ln_q\Xi_q^{(3)} + \mu\bar{N}^{(3)}. \tag{60}$$

Based on Eqs.(42) and (57), we have the following statistical expressions,

$$U_q^{(3)} = -\frac{\partial}{\partial\beta}\ln_q\Xi_q^{(3)}, \quad \bar{N}^{(3)} = -\frac{\partial}{\partial\alpha}\ln_q\Xi_q^{(3)}. \tag{61}$$

In the representation of Eq. (42), we can furthermore define enthalpy, Gibbs function and grand thermodynamic potential, respectively, as

$$H_q^{(3)} = U_q^{(3)} + PV, \tag{62}$$

$$G_q^{(3)} = U_q^{(3)} - TS_q + PV, \tag{63}$$

$$J_q^{(3)} = F_q^{(3)} - \mu\bar{N}^{(3)} = -kT\ln_q\Xi_q^{(3)}. \tag{64}$$

In next section, we study the nonextensive thermodynamics with the dual physical interpretations of thermodynamic quantities and relations.

**4 The nonextensive thermodynamic formalism with dual interpretations**

Now let us establish the thermodynamic formalism, in which the dual physical interpretations of thermodynamic quantities and corresponding thermodynamic relations are taken into account.

Without loss of the generality, we perform the study on the basis of the results deduced in the grand canonical ensemble.

We firstly introduce the nonextensive relation of Tsallis entropy,

$$S_{q1,2} = S_{q1} + S_{q2} + \frac{1-q}{k} S_{q1} S_{q2}, \tag{65}$$

where the subscripts 1 and 2 represent the observed system and the reservoir. In view of the Tsallis factor (41), the variation of Eq. (65) is

$$\delta S_{q1,2} = c_{q2} \delta S_{q1} + c_{q1} \delta S_{q2}. \tag{66}$$

According to Eq. (42), variations of the internal energy and average particle number are

$$\delta U_{q1,2}^{(3)} = \delta U_{q1}^{(3)} + \delta U_{q2}^{(3)}, \quad \delta \bar{N}_{1,2}^{(3)} = \delta \bar{N}_1^{(3)} + \delta \bar{N}_2^{(3)}. \tag{67}$$

We assume the variation of volume [37] satisfies

$$\delta V_{1,2} = c_{q2} \delta V_1 + c_{q1} \delta V_2. \tag{68}$$

When an isolated composite system arrives at the "$q$-equilibrium" state (statistical equilibrium), there should be

$$\delta S_{q1,2} = \delta U_{q1,2}^{(3)} = \delta \bar{N}_{1,2}^{(3)} = \delta V_{1,2} = 0. \tag{69}$$

By use the thermodynamic equation (53), one finds

$$\delta S_{q1,2} = c_{q2} \frac{\delta U_{q1}^{(3)} + P_1 \delta V_1 - \mu_1 \delta \bar{N}_1^{(3)}}{T_1} + c_{q1} \frac{\delta U_{q2}^{(3)} + P_2 \delta V_2 - \mu_2 \delta \bar{N}_2^{(3)}}{T_2}$$
$$= (\frac{c_{q2}}{T_1} - \frac{c_{q1}}{T_2}) \delta U_{q1}^{(3)} + c_{q2} (\frac{P_1}{T_1} - \frac{P_2}{T_2}) \delta V_1 - (\frac{c_{q2} \mu_1}{T_1} - \frac{c_{q1} \mu_2}{T_2}) \delta \bar{N}_1^{(3)} = 0. \tag{70}$$

Now that the variations in Eq. (70) are indefinite, the balance conditions can be expressed as

$$c_{q1} T_1 = c_{q2} T_2, \quad c_{q1} P_1 = c_{q2} P_2, \quad \mu_1 = \mu_2. \tag{71}$$

However, it is very strange that the thermal balance condition is marked by product of the temperature as inverse of the Lagrange multiplier and Tsallis factor. It is also strange for the mechanical balance condition. This situation is derived from such a fact that most of realistic complex systems, such as plasmas and gravitational systems, are ordinarily at nonequilibrium steady state. Therefore, if we admit the physical realities of $T$ and $P$ in Eq. (71), we cannot define an appropriate and universal temperature and pressure which mark the "local" balance, so that the Tsallis factor related to the whole (global) system appears in the definitions. On the other hand, if we regard the products in Eq. (71) as a single concept, we should introduce the following new symbols, namely,

$$T_q = c_q T, \quad P_q = c_q P. \tag{72}$$

As the single concept to identify the $q$-equilibrium state, the two concepts, on left sides of equal signs of Eq. (72), should be endowed with physical realities, yet causing at least two consequences. The first one is that the physical realities of $T$ and $P$ would be cancelled meanwhile. The second one is that such concepts defined in Eq. (72) cannot be applicable, now that most of realistic systems are in nonequilibrium states.

This conflict had troubled people for many years as an obstacle to establish a complete thermodynamic formalism. To resolve this problem, the only choice seems to be that we must simultaneously admit the physical realities of the quantities appearing in the definitions (72). That is, we must endow dual physical interpretations with the temperature, pressure and even other thermodynamic quantities and thermodynamic relations.

To simultaneously define two sets of parallel quantities in a theoretic formalism, we must put them on different but interrelated physical foundations. For this aim, let us introduce new definitions of the internal energy and average particle number, which employ the second choice of average definition [22] in NSM, namely,

$$U_q^{(2)} = \sum_{N=0}^{\infty}\sum_{s} \rho_{Ns}'^{q} E_s, \quad \overline{N}^{(2)} = \sum_{N=0}^{\infty}\sum_{s} \rho_{Ns}'^{q} N_s. \tag{73}$$

The underlying grand $q$-distribution function in Eq. (73) should be

$$\rho_{Ns}' = \frac{1}{\Xi_q^{(2)}}[1-(1-q)(\alpha' N_s + \beta' E_s)]^{\frac{1}{1-q}}, \tag{74}$$

where the grand partition function is defined as

$$\Xi_q^{(2)} \equiv \sum_{N=0}^{\infty}\sum_{s}[1-(1-q)(\alpha' N_s + \beta' E_s)]^{\frac{1}{1-q}}. \tag{75}$$

The $q$-distribution function (74) could be easily deduced by use of the maximization entropy procedure with the constraints (73) and the normalization condition. In the statistical ensemble, this $q$-distribution function in Eq. (74) is an assumption for the moment.

The composition rule for microstates of the system is temporally assumed to be

$$\alpha' N_{ij(1,2)} + \beta' E_{ij(1,2)}$$
$$= \alpha' N_{i1} + \beta' E_{i1} + \alpha' N_{j2} + \beta' E_{j2} - (1-q)[\alpha' N_{i1} + \beta' E_{i1}][\alpha' N_{j2} + \beta' E_{j2}], \tag{76}$$

where the subscripts 1 and 2 still denote the different subsystems, and the "$i$" and "$j$" represent different microstates respectively. In above rule (76), the generalized Lagrange multipliers have been introduced. It can be verified that the rule (76) assures the probability independence in the reference of distribution (74). Furthermore, one can find that in view of the rule (76), the internal energy and the average particle number in Eq.(73) satisfy

$$\alpha' \overline{N}_{1,2}^{(2)} + \beta' U_{1,2}^{(2)} = c_{q2}(\alpha' \overline{N}_1^{(2)} + \beta' U_1^{(2)}) + c_{q1}(\alpha' \overline{N}_2^{(2)} + \beta' U_2^{(2)})$$
$$-(1-q)(\alpha' \overline{N}_1^{(2)} + \beta' U_1^{(2)})(\alpha' \overline{N}_2^{(2)} + \beta' U_2^{(2)}). \tag{77}$$

The above result shows the nonadditive characters of the internal energy and average particle number in Eq. (73).

Now let us construct the fundamental thermodynamic equation in representation of Eq. (73). As the start, we introduce the following generalized Lagrange relations,

$$\beta' = \frac{1}{kT_q'}, \quad \alpha' = -\mu\beta' = -\frac{\mu}{kT_q'}, \tag{78}$$

where the chemical potential is identical to that in Eq. (44). According to Eq. (73), one gets

$$dU_q^{(2)} = q\sum_{N=0}^{\infty}\sum_{s} \rho_{Ns}'^{q-1} E_s d\rho_{Ns}' + \sum_{N=0}^{\infty}\sum_{s} \rho_{Ns}'^{q} dE_s, \tag{79}$$

$$d\overline{N}^{(2)} = q\sum_{N=0}^{\infty}\sum_{s} \rho_{Ns}'^{q-1} N_s d\rho_{Ns}' + \sum_{N=0}^{\infty}\sum_{s} \rho_{Ns}'^{q} dN_s. \tag{80}$$

Combining Eqs. (79) and (80) with the chemical potential, one gets

$$dU_q^{(2)} - \mu d\bar{N}^{(2)} = q\sum_{N=0}^{\infty}\sum_s \rho'^{q-1}_{Ns}(E_s - \mu N_s)d\rho'_{Ns} + \sum_{N=0}^{\infty}\sum_s \rho'^{q}_{Ns}d(E_s - \mu N_s). \tag{81}$$

Similarly, the first term on the right-hand side of the above equation is heat absorbed by the system from the reservoir,

$$dQ^{(2)} = q\sum_{N=0}^{\infty}\sum_s \rho'^{q-1}_{Ns}(E_s - \mu N_s)d\rho'_{Ns}$$

$$= \frac{qkT'_q[\Xi_q^{(2)}]^{1-q}}{1-q}\sum_{N=0}^{\infty}\sum_s [1-(1-q)\frac{E_s - \mu N_s}{kT'_q}]^{-1}d\rho'_{Ns}. \tag{82}$$

On the other hand, in light of Eqs.(74) and (75), Tsallis entropy satisfies

$$dS_q = \frac{qk[\Xi_q^{(2)}]^{1-q}}{1-q}\sum_{N=0}^{\infty}\sum_s [1-(1-q)\frac{E_s - \mu N_s}{kT'_q}]^{-1}d\rho'_{Ns}. \tag{83}$$

Therefore, there is

$$dQ^{(2)} = T'_q dS_q. \tag{84}$$

Furthermore, the second term on the right side of Eq.(81) is the work done by the reservoir, namely,

$$dW^{(2)} = \sum_{N=0}^{\infty}\sum_s \rho'^{q}_{Ns}d(E_s - \mu N_s) = -P'_q dV. \tag{85}$$

Substituting Eqs. (84) and (85) into Eq. (81), we immediately have,

$$dU_q^{(2)} = T'_q dS_q - P'_q dV + \mu d\bar{N}^{(2)}. \tag{86}$$

From Eq. (86), the following relations can be found,

$$T'_q = \left(\frac{\partial U_q^{(2)}}{\partial S_q}\right)_{V,\bar{N}^{(2)}}, \quad -P'_q = \left(\frac{\partial U_q^{(2)}}{\partial V}\right)_{S_q,\bar{N}^{(2)}}, \quad \mu = \left(\frac{\partial U_q^{(2)}}{\partial \bar{N}^{(2)}}\right)_{S_q,V}. \tag{87}$$

In order to define appropriate physical quantities which can perfectly characterize the balance conditions, let us consider variation of Eq. (77), which would lead to,

$$\delta(\alpha'\bar{N}_{1,2}^{(2)} + \beta' U_{1,2}^{(2)}) = c_{q2}\delta(\alpha'\bar{N}_1^{(2)} + \beta' U_1^{(2)}) + c_{q1}\delta(\alpha'\bar{N}_2^{(2)} + \beta' U_2^{(2)}). \tag{88}$$

Consider the indefiniteness of the generalized Lagrange multipliers, we further have

$$\delta\bar{N}_{1,2}^{(2)} = c_{q2}\delta\bar{N}_1^{(2)} + c_{q1}\delta\bar{N}_2^{(2)}, \quad \delta U_{1,2}^{(2)} = c_{q2}\delta U_1^{(2)} + c_{q1}\delta U_2^{(2)}. \tag{89}$$

Then, in view of Eqs. (66), (68), (86) and (89), the $q$-equilibrium state should obey,

$$\delta S_{q1,2} = c_{q2}\frac{\delta U_{q1}^{(2)} + P'_{q1}\delta V_1 - \mu_1\delta\bar{N}_1^{(2)}}{T'_{q1}} + c_{q1}\frac{\delta U_{q2}^{(2)} + P'_{q2}\delta V_2 - \mu_2\delta\bar{N}_2^{(2)}}{T'_{q2}}$$

$$= c_{q2}(\frac{1}{T'_{q1}} - \frac{1}{T'_{q2}})\delta U_{q1}^{(2)} + c_{q2}(\frac{P'_{q1}}{T'_{q1}} - \frac{P'_{q2}}{T'_{q2}})\delta V_1 - c_{q2}(\frac{\mu_1}{T'_{q1}} - \frac{\mu_2}{T'_{q2}})\delta\bar{N}_1^{(2)} = 0. \tag{90}$$

Now that the variations in Eq. (90) are indefinite, the balance conditions can be expressed as

$$T'_{q1} = T'_{q2}, \quad P'_{q1} = P'_{q2}, \quad \mu_1 = \mu_2. \tag{91}$$

We can see that the balance conditions are concise in mathematical forms. Besides, it is reasonable to assume that the balances conditions (71) and (91) describe the same $q$-equilibrium state. So, we have that

$$T'_q = T_q = c_q T, \quad P'_{q1} = P_{q2} = c_q P. \tag{92}$$

The first expression in Eq.(92) is the temperature duality [29], which simultaneously confirms the physical realities of the temperature $T$ defined in Eq.(54) and the temperature $T_q$ defined in Eq. (87). In order to distinguish them, we call the former one Lagrange temperature, and call the latter one physical temperature, which is firstly proposed by Abe in [25]. Accordingly, we call the pressure in Eq. (54) Lagrange pressure, and call the pressure in Eq. (87) physical pressure. In one word, the quantities defined in the representation (42) are called the Lagrange ones and the thermodynamic quantities defined in the representation (73) are called the physical ones.

Based on Eq. (92), we introduce the following relations

$$\alpha' = \frac{\alpha}{c_q}, \quad \beta' = \frac{\beta}{c_q}, \quad \Xi_q^{(2)} = \Xi_q^{(3)}, \tag{93}$$

according to Eq. (93), the grand canonical $q$-distribution function (74) is completely identical to that in Eq.(57). Also according to the relations in Eq.(93), the composition rule (76) can be directly derived from the rule (29) by simply setting the internal energy and average particle number to be zeros. On the other hand, in view of Eq.(92), from Eqs.(51) and (84) one can obtain that

$$dQ^{(2)} = c_q dQ^{(3)}. \tag{94}$$

Besides, from Eqs.(52) and (85) one can get that

$$dW^{(2)} = c_q dW^{(3)}. \tag{95}$$

Then we have

$$dU_q^{(2)} - \mu d\bar{N}^{(2)} = c_q (dU_q^{(3)} - \mu d\bar{N}^{(3)}). \tag{96}$$

Now that the chemical potential is indefinite, one has

$$dU_q^{(2)} = c_q dU_q^{(3)}, \quad d\bar{N}^{(2)} = c_q d\bar{N}^{(3)}. \tag{97}$$

Furthermore, based on Eqs.(42) and (73), we get the links,

$$U_q^{(2)} = c_q U_q^{(3)}, \quad \bar{N}^{(2)} = c_q \bar{N}^{(3)}. \tag{98}$$

Therefore, in view of Eq.(93) and from Eq.(59) we can directly obtain

$$c_q = [\Xi_q^{(2)}]^{1-q} + (1-q)\frac{U_q^{(2)} - \mu \bar{N}^{(2)}}{kT_q}, \tag{99}$$

which can also be calculated through Eqs.(73) and (74). And the following statistical expressions are easily obtained,

$$U_q^{(2)} = -\frac{\partial}{\partial \beta'} \ln_q \Xi_q^{(2)}, \quad \bar{N}^{(2)} = -\frac{\partial}{\partial \alpha'} \ln_q \Xi_q^{(2)}. \tag{100}$$

Then the physical free energy is defined as

$$F_q^{(2)} = U_q^{(2)} - T_q S_q = -kT_q \ln_q \Xi_q^{(2)} + \mu \bar{N}^{(2)}. \tag{101}$$

It is easy to find that

$$F_q^{(2)} = c_q F_q^{(3)}. \tag{102}$$

In the representation of Eq. (73), we can further define the physical enthalpy, physical Gibbs function and the physical grand thermodynamic potential, respectively, in the nonextensive thermodymamics,

$$H_q^{(2)} = U_q^{(2)} + P_q V = c_q H_q^{(3)}, \tag{103}$$

$$G_q^{(2)} = U_q^{(2)} - T_q S_q + P_q V = c_q G_q^{(3)}, \tag{104}$$

$$J_q^{(2)} = F_q^{(2)} - \mu \bar{N}^{(2)} = -kT_q \ln_q \Xi_q^{(2)} = c_q J_q^{(3)}. \tag{105}$$

Then, a series of thermodynamic relations can be obtained through standard methods [29].

In summary, in this section we construct the nonextensive thermodynamic formalism in realm of the grand canonical ensemble, which consists of two sets of parallel Legendre transformation structures. One is the physical set and the other is the Lagrange set. They are linked through the Tsallis factor. One key question here is : what are the differences between them?

To answer this question, we should realize that the Lagrange internal energy (L-energy) is additive, while the physical internal energy (P-energy) is nonadditive. This inspires us to confirm that the L-energy can be regarded as sum of molecular kinetic energies and the short-range potential energies, and the P-energy should be thought as sum of all kinds of energies such as kinetic energies, long-rang and short-range potential energies, and even the radiation energies. Due to that the thermometer cannot feel long-range potential energy, the physical temperature related to P-energy is unable to measure in an experiment, while the Lagrange temperature associated with L-energy is able to measure experimentally.

Therefore, the Lagrange temperature is identical to the temperature defined in the traditional thermodynamics, to some extent. The Lagrange heat related to Lagrange temperature is measured in a local Maxiweillian equilibrium. The Lagrange work can be explained in the same sense; then the Lagrange pressure is also measured in the local Maxiweillian equilibrium sense. In contrast to this, the physical heat includes the radiation heat and the physical work related to the variation of system configuration. They are both unable to measure directly.

The Lagrange average particle number (L-number) only includes the noncreative particles inside the system, and the physical average particle number (P-number) includes the creative particles and the those taking part in the interactions between the system and the reservoir.

In one word, the quantities in Lagrange set of Legendre transform formalism are related to the local physical processes and therefore are easy to measure. The quantities in the physical set are mainly related to the nonlocal physical processes, so they are uneasy to measure. However, the quantities in the physical set can be obtained through those in the Lagrange set, provided the Tsallis factor is given.

**5 The single-particle *q*-distribution in the canonical ensemble**

Boltzmann statistics has a single-particle distribution function, whose normalization is made in the 6-dimension phase space, where the volume element is written as $d\tau = d^3\vec{v}d^3\vec{r}$. Now let us deduce the nonextensive version of the Boltzmann single-particle distribution function in the realm of canonical ensemble. In this section, we employ the canonical *q*-distribution function (25).

For a closed system at quantum state *s*, its energy can be explained as sum of all the single-particle energies, that is, $E_s = \sum_{i=1}^{N} \varepsilon_i$. Now that the energy distributions for the single particles are identical, we can directly assume that $E_s = N\varepsilon_1$. According to this assumption, the canonical partition function can be expressed as

$$Z_q = \int [1-(1-q)\beta'N\varepsilon_1]^{\frac{1}{1-q}} d\tau_1 \cdots d\tau_N. \tag{106}$$

In order to obtain the single-particle $q$-distribution function, we introduce the following parameter transformation,

$$(1-q)N = 1-\nu, \tag{107}$$

where a new nonextensive parameter $\nu$ is given. In fact, this transformation constructs a link between statistical ensemble and molecular dynamic system, and endows different physical senses to these two types of parameters. Their values would be evaluated later. In light of the parameter transformation (107), one can easily find that

$$\begin{aligned} Z_q &= \int [1-(1-\nu)\beta'\varepsilon_1]^{\frac{N}{1-\nu}} d\tau_1 \cdots d\tau_N \\ &= \prod_{i=1}^{N} \int [1-(1-\nu)\beta'\varepsilon_1]^{\frac{1}{1-\nu}} d\tau_1 = Z_{\nu 1}^N, \end{aligned} \tag{108}$$

where the single-particle partition function is defined as

$$Z_{\nu 1} = \int [1-(1-\nu)\beta'\varepsilon_1]^{\frac{1}{1-\nu}} d\tau_1. \tag{109}$$

Then the single-particle $\nu$-distribution function is

$$f = \frac{1}{Z_{\nu 1}}[1-(1-\nu)\beta'\varepsilon_1]^{\frac{1}{1-\nu}}, \tag{110}$$

and the canonical $q$-distribution function can be written as the $N$ power of the $\nu$-distribution,

$$\rho_s = \frac{1}{Z_{\nu 1}^N}[1-(1-\nu)\beta'\varepsilon_1]^{\frac{N}{1-\nu}} = f^N. \tag{111}$$

The single-particle Tsallis factor can be given as

$$c_\nu = \int f^\nu d\tau = \frac{1}{Z_{\nu 1}^\nu} \int [1-(1-\nu)\beta'\varepsilon_1]^{\frac{\nu}{1-\nu}} d\tau_1, \tag{112}$$

from which it is easily found that

$$\begin{aligned} c_q &= \sum \rho_s^q = \frac{1}{Z_q^q} \int [1-(1-q)\beta'N\varepsilon_1]^{\frac{q}{1-q}} d\tau_1 \cdots d\tau_N \\ &= \frac{1}{Z_{\nu 1}^{N-1+\nu}} \int [1-(1-\nu)\beta'\varepsilon_1]^{\frac{N-1+\nu}{1-\nu}} d\tau_1 \cdots d\tau_N \\ &= \frac{1}{Z_{\nu 1}^\nu} \int [1-(1-\nu)\beta'\varepsilon_1]^{\frac{\nu}{1-\nu}} d\tau_1 = c_\nu. \end{aligned} \tag{113}$$

That is, the Tsallis factor of the system is identical to the single-particle Tsallis factor.

The L-energy of the system is given by

$$\begin{aligned} U_q^{(3)} &= \frac{\sum \rho_s^q E_s}{\sum \rho_s^q} = N \frac{\int \varepsilon_1 [1-(1-q)\beta'N\varepsilon_1]^{\frac{q}{1-q}} d\tau_1 \cdots d\tau_N}{\int [1-(1-q)\beta'N\varepsilon_1]^{\frac{q}{1-q}} d\tau_1 \cdots d\tau_N} \\ &= N \frac{\int \varepsilon_1 [1-(1-\nu)\beta'\varepsilon_1]^{\frac{\nu}{1-\nu}} d\tau_1}{\int [1-(1-\nu)\beta'\varepsilon_1]^{\frac{\nu}{1-\nu}} d\tau_1} = N \frac{\int \varepsilon_1 f^\nu d\tau_1}{\int f^\nu d\tau_1}. \end{aligned} \tag{114}$$

In the L-set of thermodynamic formalism, if we define the single-particle averaged energy as

$$\overline{\varepsilon}^{(3)} = \frac{U_q^{(3)}}{N}, \tag{115}$$

we see that

$$\overline{\varepsilon}^{(3)} = \frac{\int \varepsilon f^\nu d\tau}{\int f^\nu d\tau} = -\frac{\partial}{\partial \beta}\ln_\nu Z_\nu \equiv -\frac{\partial}{\partial \beta}\frac{Z_\nu^{1-\nu}-1}{1-\nu}, \tag{116}$$

where the subscript 1 has been omitted. It is apparent that the above calculation method for the single-particle averaged energy has been applied popularly [36].

Similarly, the P-energy of the system is given by

$$U_q^{(2)} = \sum \rho_s^q E_s = \frac{N}{Z_q^q}\int \varepsilon_1[1-(1-q)\beta' N\varepsilon_1]^{\frac{q}{1-q}} d\tau_1 \cdots d\tau_N$$

$$= \frac{N}{Z_{\nu 1}^{N-1+\nu}}\int \varepsilon_1[1-(1-\nu)\beta'\varepsilon_1]^{\frac{N-1+\nu}{1-\nu}} d\tau_1 \cdots d\tau_N$$

$$= \frac{N}{Z_{\nu 1}^\nu}\int \varepsilon_1[1-(1-\nu)\beta'\varepsilon_1]^{\frac{\nu}{1-\nu}} d\tau_1 = N\int \varepsilon_1 f^\nu d\tau_1. \tag{117}$$

We can also define the single-particle $q$-averaged energy as, in the P-set of formalism,

$$\overline{\varepsilon}^{(2)} = \frac{U_q^{(2)}}{N}. \tag{118}$$

Then one has

$$\overline{\varepsilon}^{(2)} = \int \varepsilon f^\nu d\tau = -\frac{\partial}{\partial \beta'}\ln_\nu Z_\nu, \quad \overline{\varepsilon}^{(2)} = c_\nu \overline{\varepsilon}^{(3)}. \tag{119}$$

Considering the single-particle distribution function (110), and in view of Eqs. (107) and (111), the $q$-entropy of the canonical ensemble becomes

$$S_q = Nk\frac{\int f_1^{qN} d\tau_1 \cdots d\tau_N - 1}{1-\nu} = Nk\frac{\int f_1^{N-1+\nu} d\tau_1 \cdots d\tau_N - 1}{1-\nu}$$

$$= Nk\frac{\int f_1^\nu d\tau_1 - 1}{1-\nu} = NS_\nu, \tag{120}$$

where the single particle $\nu$-entropy is defined as

$$S_\nu \equiv k\frac{\int f^\nu d\tau - 1}{1-\nu}. \tag{121}$$

In the kinetics, the normalization of the single-particle distribution function is actually equal to the particle number, so we modify the distribution, omitting the subscript, as

$$f = \frac{N}{Z_\nu}[1-(1-\nu)\frac{\varepsilon}{kT_\nu}]^{\frac{1}{1-\nu}}, \tag{122}$$

where $T_\nu$ is the physical temperature, ao that Eq.(113) holds. The function (122) is still not the familiar mathematical expression in the nonextensive kinetics [39], in which the temperature should be explained as the Lagrange temperature $T$. In order to explain this, let us apply Eq.(122) to a self-gravitating gaseous system, where the physical temperature is equivalent to the concept of the gravitational temperature $T_g$, given [40] by

$$kT_g = kT + (1-\nu)m(\varphi - \varphi_0), \tag{123}$$

where $\varphi$ is the gravitational potential. The gravitational temperature should be constant in a whole self-gravitating system for the *q*-equilibrium state. Meanwhile, the particle energy in Eq.(122) is explained as

$$\varepsilon = \frac{1}{2}mv^2 + m(\varphi - \varphi_0).\qquad(124)$$

Substituting Eqs.(123) and (124) into Eq.(122), we can find that

$$f = \frac{N}{Z_\nu}[1-(1-\nu)\frac{m(\varphi-\varphi_0)}{kT_\nu}]^{\frac{1}{1-\nu}}[1-(1-\nu)\frac{\frac{1}{2}mv^2}{kT}]^{\frac{1}{1-\nu}},\qquad(125)$$

which now is familiar to us in Refs.[5, 39-41].

Now let us evaluate the values of these two types of nonextensive parameters in Eq.(107). As everyone knows, the nonextensivity for a nonextensive system is measured by the value of the parameter (1-*q*). According to Eq.(107), the value of (1-*q*) can be determined by the particle number and the value of new parameter (1-*v*). In order to evaluate the new parameter, let us consider some abnormal states of the nonextensive systems. For instance, in the critical state of gas-liquid phase, there are

$$\left(\frac{\partial P}{\partial V}\right)_T = 0, \quad \left(\frac{\partial^2 P}{\partial V^2}\right)_T = 0.\qquad(126)$$

On the other hand, the L-pressure for the nonextensive ideal gas [29] is

$$PV = NkTZ_q^{1-q},\qquad(127)$$

with the partition function,

$$Z_q = a(N,q)V^N(kT_q)^{\frac{3}{2}N},\qquad(128)$$

where the quantity *a*(*N*, *q*) has a certain expression [25, 29]. It can be found that

$$\left(\frac{\partial P}{\partial V}\right)_T = \frac{NkTZ_q^{1-q}}{V^2}[N(1-q)-1]=0,\qquad(129)$$

$$\text{and } \left(\frac{\partial^2 P}{\partial V^2}\right)_T = \frac{NkTZ_q^{1-q}}{V^3}[N(1-q)-1][N(1-q)-2]=0,\qquad(130)$$

from which we obtain

$$N(1-q) = 1-\nu = 1.\qquad(131)$$

According to this result, now that the value of (1-*v*) is fixed, we conclude that in this case the value of (1-*q*) is inversely proportional to particle number of system. Furthermore, in gaseous self-gravitating systems, where the long-range interactions can be thought as the manifestation of some kind of "phase transition", the typical value of (1-*v*) is approximately 0.3 [42]. Therefore, in the gaseous self-gravitating systems, the value of (1-*q*) is also inversely proportional to the particle number. Through the pure ensemble discussion [43], one can obtain the relation similar to Eq.(131), showing that the value of (1-*q*) is limited by the particle number of system.

However, what should be emphasized here is that expression (131) holds only in the case of gas-liquid phase transition. In normal cases for nonextensive systems, the value of (1-*v*) could be much less than unity. For instance, only with such an approximation, (1-*v*)→+0, the calculated result of the second viral coefficient of a nonextensive gas fits very well the experimental curve [44]. Therefore, we can confirm that in normal systems the value of *N*(1-*q*) is small, indicating that in limit of large particle number, value of the parameter (1-*q*) is very tiny, leading to those approximations of Eqs.(22) and (55).

**6 The nonextensive quantum statistics with the parameter transformation**

By using the Green function method [45], the nonextensive quantum statistical formulae can be exactly calculated for Bosons and Fermions. However, the mathematical forms are complicated and their applications to realistic systems are difficult. In the dilute gas approximation [44,45], the expressions for the nonextensive quantum statistics seem simple and familiar, yet it is limited also in its applications now that this approximation only holds in high temperature case. In this section, we use the parameter transformation to study the nonextensive quantum statistics, in which the obtained expressions are similar to those using the dilute gas approximation, yet with more wide applicability. We employ the grand $q$-distribution function (57) or (74) here.

Firstly, let us consider a system consisting of one kind of particles, whose energy level is marked by $\varepsilon_l$ ($l=1,2,...$). Assuming all the energy levels are non-degenerate, the particle number and energy can be written as

$$N = \sum_l a_l, \quad E = \sum_l \varepsilon_l a_l, \tag{132}$$

where $\{a_l\}$ denote the particle distribution of the considered system on each energy level $l$. In the grand canonical ensemble, the distribution $\{a_l\}$ is indefinite; so we must consider all the possibilities in the grand canonical sum. This means the grand partition function to be

$$\Xi_q = \sum_{\{a_l\}} [1-(1-q)\sum_l (\alpha' + \beta' \varepsilon_l) a_l]^{\frac{1}{1-q}}. \tag{133}$$

Now that the distribution $\{a_l\}$ is not uniform on each energy level, it is difficult to propose the parameter transformation like Eq.(107); however, we can let

$$[1-(1-q)\sum_l (\alpha' + \beta' \varepsilon_l) a_l]^{\frac{1}{1-q}} = \prod_l [1-(1-\nu_l)(\alpha' + \beta' \varepsilon_l) a_l]^{\frac{1}{1-\nu_l}}$$

$$\approx [1-(1-\nu_l)(\alpha' + \beta' \varepsilon_l) a_l]^{\frac{N_l}{1-\nu_l}}, \tag{134}$$

where the following parameter transformation is employed,

$$(1-q)N_l = 1-\nu_l, \tag{135}$$

where the quantity $N_l$ is an effective number related to the possible number of energy level for given particle number distribution. The nonextensive parameter $\nu_l$ is related to one given energy level. In view of Eq.(134), the grand partition function (133) becomes

$$\Xi_q = \prod_l \sum_{a_l} [1-(1-\nu_l)(\alpha' + \beta' \varepsilon_l) a_l]^{\frac{1}{1-\nu_l}} \equiv \prod_l \Xi_l \approx \Xi_l^{N_l}, \tag{136}$$

where the $\Xi_l$ is the partition function in given energy level. Then the distribution function in given energy level, or subsystem with energy level $\varepsilon_l$ can be given by

$$\rho_l = \frac{1}{\Xi_l}[1-(1-\nu_l)(\alpha' + \beta' \varepsilon_l) a_l]^{\frac{1}{1-\nu_l}}. \tag{137}$$

Then the averaged particle number in P-set of formalism is

$$\bar{a}_l^{(2)} = \sum_{Ns} a_l \rho_{Ns}^q$$

$$= \frac{1}{\Xi_l^{N_l-1+\nu_l}} \sum_{a_l} a_l [1-(1-\nu_l)(\alpha' + \beta' \varepsilon_l) a_l]^{\frac{\nu_l}{1-\nu_l}} \prod_{m\neq l} \sum_{a_m} [1-(1-\nu_l)(\alpha' + \beta' \varepsilon_m) a_m]^{\frac{1}{1-\nu_l}}$$

$$= \frac{1}{\Xi_l^{\nu_l}} \sum_{a_l} a_l [1-(1-\nu_l)(\alpha' + \beta' \varepsilon_l) a_l]^{\frac{\nu_l}{1-\nu_l}} = \sum_{a_l} a_l \rho_l^{\nu_l}. \tag{138}$$

It is easy to find that

$$c_q \equiv \sum_{Ns} \rho_{Ns}^q$$

$$= \frac{1}{\Xi_l^{N_l-1+v_l}} \sum_{a_l}[1-(1-v_l)(\alpha'+\beta'\varepsilon_l)a_l]^{\frac{v_l}{1-v_l}} \prod_{m \neq l}\sum_{a_m}[1-(1-v_l)(\alpha'+\beta'\varepsilon_m)a_m]^{\frac{1}{1-v_l}}$$

$$= \frac{1}{\Xi_l^{v_l}}\sum_{a_l}[1-(1-v_l)(\alpha'+\beta'\varepsilon_l)a_l]^{\frac{v_l}{1-v_l}} = \sum \rho_l^{v_l} \equiv c_{v_l}, \qquad (140)$$

which shows equivalence between the Tsallis factor of the system and the Tsallis factor of the subsystem. Likewise, for the averaged particle number in L-set of formalism, there is

$$\bar{a}_l^{(3)} = \frac{\sum_{Ns} a_l \rho_{Ns}^q}{\sum_{Ns} \rho_{Ns}^q} = \frac{\sum_{a_l} a_l [1-(1-v_l)(\alpha'+\beta'\varepsilon_l)a_l]^{\frac{v_l}{1-v_l}}}{\sum_{a_l}[1-(1-v_l)(\alpha'+\beta'\varepsilon_l)a_l]^{\frac{v_l}{1-v_l}}}$$

$$= \frac{\sum a_l \rho_l^{v_l}}{\sum \rho_l^{v_l}} = \frac{\bar{a}_l^{(2)}}{c_{v_l}}. \qquad (141)$$

Now we employ the second parameter transformation, namely,

$$(1-v_l)a_l = 1-v. \qquad (142)$$

This transformation (142) further merges the nonextensive parameter $v_l$ and the particle number in given energy level. And one assumption has been adopted that the value of $v$ is irrelevant of the energy level. Combing this transformation (142) with Eq.(135), we arrive at

$$(1-q)N = 1-v, \qquad (143)$$

which is identical to Eq.(107) in the canonical ensemble, and $N$ here is explained as the (average) particle number of the quantum system. Just as mentioned in last section, the value of (1-$v$) for most of the nonextensive systems is small.

We have confirmed that the quantum effect can suppress the nonextensive effect to some extent, which further reduces the value of (1-$v$). The fundamental reason might be that, due to the uncertainty principle the position and size of the quantum cannot be measured. This is equal to say, the quantum statistics is approximately irrelevant to the physical processes or phenomena where the position and size of particles are important. On the contrary, the nonexetnsive effect is relevant to the position of particles, such as the systems including long-range interactions and correlations [5,40, 41], or the sizes of particles, such as the sizes of molecules or the Van der Waals radii [46]. About the second case, let us make more interpretation. In Ref [46], the author obtained one simple expression (see Eq. (35) in Ref [46]) in the nonextensive gas system, which relates to the nonextensive parameter (1-$v$) to a combined quantity $\lambda$, which is proportional to the third power of molecular size or the Van der Waals radius. It can be found that when the molecular size tends to zero, the parameter (1-$v$) also tends to zero.

Therefore, now that the size of quantum particle can be regarded as to be arbitrarily small, we can conclude that the value of $v$ in nonextensive quantum statistics is very close to the unity, namely, $v \to 1$.

According to this, in the nonextensive quantum statistics we can adopt

$$c_{v_l} \equiv \sum \rho_l^{v_l} \approx 1. \qquad (144)$$

On basis of Eq. (144), the averaged particle number in nonextensive quantum statistics is derived by

$$\bar{a}_l = \bar{a}_l^{(3)} = \frac{\sum_{a_l} a_l [1-(1-\nu)(\alpha+\beta\varepsilon_l)]^{\frac{a_l-1+\nu}{1-\nu}}}{\sum_{a_l}[1-(1-\nu)(\alpha+\beta\varepsilon_l)]^{\frac{a_l-1+\nu}{1-\nu}}}$$

$$= \frac{\sum_{a_l} a_l [1-(1-\nu)(\alpha+\beta\varepsilon_l)]^{\frac{a_l-1}{1-\nu}}}{\sum_{a_l}[1-(1-\nu)(\alpha+\beta\varepsilon_l)]^{\frac{a_l-1}{1-\nu}}}, \quad (145)$$

where the parameter transformation (142) is considered and the difference between the L-temperature and the P-temperature is ignored. For convenience, we define the quantity that

$$t \equiv [1-(1-\nu)(\alpha+\beta\varepsilon_l)]^{\frac{1}{1-\nu}}, \quad (146)$$

which always satisfies $0 < t < 1$, no matter what the value of $(1-\nu)$ is. Then for the nonextensive Fermions, based on Eq.(145) we have

$$\bar{a}_{lF} = \frac{1}{t^{-1}+1} = \frac{1}{[1+(\nu-1)(\alpha+\beta\varepsilon_l)]^{\frac{1}{\nu-1}}+1}, \quad (147)$$

and for the nonextensive Bosons, we get

$$\bar{a}_{lB} = \frac{\sum_{a_l=0}^{\infty} a_l t^{a_l-1}}{t^{-1}\sum_{a_l=0}^{\infty} t^{a_l}} = \frac{\frac{\partial}{\partial t}\sum_{a_l=0}^{\infty} t^{a_l}}{\frac{t^{-1}}{1-t}} = t(1-t)\frac{\partial}{\partial t}\left(\frac{1}{1-t}\right) = \frac{t}{1-t}$$

$$= \frac{1}{[1+(\nu-1)(\alpha+\beta\varepsilon_l)]^{\frac{1}{\nu-1}}-1}. \quad (148)$$

It is obvious that the above two expressions (147) and (148) for nonextensive quantum statistics are equivalent to those calculated in the dilute gas approximation [47,48]. However, the parameter transformation adopted here is obviously more accurate and the divergence problem [47] at low temperature regions is also avoided.

**7 Conclusions and discussions**

In this work, on basis of the equal probability principle, we study the statistical ensembles in the framework of NSM, where the complex systems containing long-range interactions and long-range correlations are investigated. In the canonical ensemble, Tsallis entropy can be regarded as a function of energy fluctuation, and the expansion of Tsallis entropy easily produces the generalized Gibbs distribution function with power-law $q$-form in Eq.(11). And in the grand canonical ensemble, by also making the Taylor expansion of the reservoir entropy as a function of energy and particle number fluctuations, we can get the generalized grand distribution function also with power-law $q$-form in Eq.(38). It is interesting that the Taylor expansion employed in this paper is independent of the large reservoir assumption.

Two power-law $q$-distribution functions (11) and (38) are identical to those calculated by the maximization entropy procedure. However, due to the self-referential property, they are difficult to be applied to realistic complex systems. So they are modified as the forms only containing the system energy and particle number, see Eqs. (25) and (57), where two relations are suggested which can directly deduce the expressions of Lagrange free energy in both canonical and grand canonical ensembles.

On basis of Eq.(38), in representation of Eq.(42) we obtain the fundamental thermodynamic equation (53). By assuming that the volume is nonadditive, the balance conditions (11) for $q$-equilibrium state are obtained in light of Eq. (53). However, it is strange that in the balance conditions, there exists the Tsallis factor which is related to the global property of the system. The key point is that, the temperature defined in theory is not identical to that in realistic systems, now that most of the complex systems are at nonequilibrium states. This strongly suggests that we should give dual interpretations to the thermodynamic quantities such as the temperature, pressure and internal energy, and thermodynamic relations. In order to carry out this, we propose a new definition of average in Eq.(73). And then the other thermodynamic equation (86) is derived, on basis of which new balance conditions (91) are presented, where new quantities such the physical pressure and the physical temperature are defined.

Then we establish the nonextensive thermodynamic formalism in the statistical ensemble, which consists of two parallel Legendre transformation structures. Among them one is the Lagrange set and the other is the physical set. The quantities in the Lagrange set are measurable in experiments, and the quantities in the physical set are unmeasurable experimentally. They are linked through the Tsallis factor.

We have developed a parameter transformation technique to construct the link between the statistical ensemble and the basic physical kinetics in NSM. By use of this transformation, the single-particle $q$-distribution function can be deduced exactly from the generalized power-law Gibbs $q$-distribution. In light of this single particle $q$-distribution, we have proved that the system entropy is product of the particle number and the single particle entropy in Eq.(120). This shows that the $q$-distribution deduced here is just the kinetic $q$-distribution function derived from the $q$-H theorem on basis of the generalized Boltzmann equation [39-41].

At the critical state of liquid-gas phase, the first and second order partial derivatives of the pressure for the volume are both zeros, from which we find that $N(1-q)=1$ in Eq.(131). This means that the value of (1-$q$) is inversely proportional to the particle number of system. The same conclusions are obtained in gaseous self-gravitating systems [42] and the normal ensemble [43]. On the other hand, for most complex systems in normal situations, the value of $N(1-q)$ tends to zero [44]. Therefore, the real value of (1-$q$) in normal states of complex systems is very tiny, guaranteeing the validity of these approximations Eqs.(22) and (55).

For nonextensive quantum statistics, the Green function method can be used to give exact results, but the quantum statistical expressions derived from this method are complicated so that it seems difficult to apply in realistic complex systems. By use of the dilute gas approximation, the familiar expressions in the nonextensive quantum statistics can be obtained. However, this approach is only appropriate for the extreme situation of very high temperature. This parameter transformation approach can overcome the two defects. By use of it, we can bypass the obstacle of factorization in the power-law $q$-distribution functions and in quantum sum of the partition function in Eq.(133), and so we can obtain the familiar forms of the Fermi and Boson $q$-distributions in an exact fashion in the nonextensive quantum statistics.

**Acknowledgements**

This work is supported by the National Natural Science Foundation of China under Grant No.

11405092, and also by National Natural Science Foundation of China under Grant No. 11775156.11405092, and also by National Natural Science Foundation of China under Grant No. 11775156.11405092, and also by National Natural Science Foundation of China under Grant No. 11775156.

**Appendix**

Ordinarily, we have that

$$\Omega_r(E_0 - E_s) = e_q^{\left[\ln_q \Omega_r(E_0 - E_s)\right]}, \tag{A.1}$$

where

$$e_q^x \equiv [1+(1-q)x]^{\frac{1}{1-q}}, \quad \ln_q x \equiv \frac{x^{1-q}-1}{1-q}. \tag{A.2}$$

Then according to Eq.(8), we obtain that

$$\Omega_r(E_0 - E_s) = e_q^{\left[\ln_q \Omega_r(U_{qr}^{(3)}) - \beta_r(E_s - U_q^{(3)})\right]}$$

$$= \left[1+(1-q)(\frac{[\Omega_r(U_{qr}^{(3)})]^{1-q}-1}{1-q}) - (1-q)\beta_r(E_s - U_q^{(3)})\right]^{\frac{1}{1-q}}$$

$$= \left[[\Omega_r(U_{qr}^{(3)})]^{1-q} - (1-q)\beta_r(E_s - U_q^{(3)})\right]^{\frac{1}{1-q}}$$

$$= \Omega_r(U_{qr}^{(3)})\left[1-(1-q)\frac{\beta_r}{c_{qr}}(E_s - U_q^{(3)})\right]^{\frac{1}{1-q}}.$$

(A.3)

where

$$c_{qr} = [\Omega_r(U_{qr}^{(3)})]^{1-q}. \tag{A.4}$$

In view of Eq.(10), we derive that

$$\rho_s = \frac{\Omega_r(E_0 - E_s)}{\Omega_t(E_0)} = \frac{\Omega_r(U_{qr}^{(3)})}{\Omega_t(E_0)}\left[1-(1-q)\frac{\beta}{c_q}(E_s - U_q^{(3)})\right]^{\frac{1}{1-q}}. \tag{A.5}$$

By defining

$$\overline{Z}_q^{(3)} = \frac{\Omega_t(E_0)}{\Omega_r(U_{qr}^{(3)})} = \sum_s [1-(1-q)\frac{\beta}{c_q}(E_s - U_q^{(3)})]^{\frac{1}{1-q}}, \tag{A.6}$$

we obtain

$$\rho_s = \frac{1}{\overline{Z}_q^{(3)}}\left[1-(1-q)\frac{\beta}{c_q}(E_s - U_q^{(3)})\right]^{\frac{1}{1-q}}. \tag{A.7}$$

This is Eq. (11).

**References**

1. Tsallis, C.: Possible generalization of Boltzmann–Gibbs statistics. J. Stat. Phys. **52**, 479 (1988)
2. Du, J.: The nonextensive parameter and Tsallis distribution for self-gravitating systems. Europhys. Lett. **67**, 893 (2004)
3. Leubner, M. P.: Nonextensive theory of dark matter and gas density profiles. Astrophys. J. **632**, L1 (2005)
4. Du, J.: The Chandrasekhar's condition of the equilibrium and stability for a star in the nonextensive kinetic theory. New Astron. **12**, 60 (2006)
5. Du, J.: Nonextensivity and the power-law distributions for the systems with self-gravitating long-range interactions. Astrophys. Space Sci. **312**, 47 (2007) and the references therein.
6. Cardone, V. F., Leubner, M. P. and Popolo, A. Del: Spherical galaxy models as equilibrium configurations in non-extensive statistics. Mon. Not. R. Astron. Soc. **414,** 2265–2274 (2011)

7. Zheng, Y. : The thermodynamic stability criterion in a self-gravitational system with phase transition. EPL **102**, 10009 (2013)
8. Zheng, Y., Luo, W., Li, Q. and Li, J.: The polytropic index and adiabatic limit: Another interpretation to the convection stability criterion. EPL **102**, 10009 (2013)
   Zheng, Y. and Du, J. : Two physical explanations of the nonextensive parameter in a self-gravitating system. EPL **107**, 60001 (2014)
9. Lima, J. A. S., Silva, Jr R., and Santos, J. : Plasma oscillations and nonextensive statistics. Phys. Rev. E, **61**, 3260 (2000)
10. Du, J. : Nonextensivity in nonequilibrium plasma systems with Coulombian long-range interactions. Phys. Lett. A **329**, 262 (2004)
11. Liu, L. and Du, J. : Ion acoustic waves in the plasma with the power-law q-distribution in nonextensive statistics. Physica A **387**, 4821 (2008)
12. Yu, H. and Du, J. : The nonextensive parameter for the rotating astrophysical systems with power-law distributions. EPL **116**, 60005 (2016)
13. Du, J.: Power-law distributions and fluctuation-dissipation relation in the stochastic dynamics of two-variable Langevin equations. J. Stat. Mech. P02006 (2012)
14. Guo, R. and Du, J.: The precise time-dependent solution of the Fokker–Planck equation with anomalous diffusion. Ann. Phys. **359**, 187 (2015)
15. Guo, R. and Du, J.: Power-law behaviors from the two-variable Langevin equation: Ito's and Stratonovich's Fokker-Planck equations. J. Stat. Mech. P02015 (2013)
16. Oikonomou, T., Provata, A. and Tirnakli, U.: Nonextensive statistical approach to non-coding human DNA. Physica A **387**, 2653 (2008)
17. Rolinski, O. J., Martin, A. and Birch, D. J. S.: Human serum albumin-flavonoid interactions monitored by means of tryptophan kinetics. Ann. New York Academy Sci. **1130**, 314 (2008)
18. Du, J.: Transition state theory: A generalization to nonequilibrium systems with power-law distributions. Physica A **391**, 1718 (2012)
19. Eftaxias, K., Minadakis, G., Potirakis, S. M., et al.: Dynamical analogy between epileptic seizures and seismogenic electromagnetic emissions by means of nonextensive statistical mechanics. Physica A **392**, 497 (2013)
20. Yin, C. and Du, J.: The power-law reaction rate coefficient for an elementary bimolecular reaction. Physica A **395**, 416 (2014)
21. Curado, E. M. F. and Tsallis, C.: Generalized statistical mechanics: connection with thermodynamics. J. Phys. A: Math. Gen., **24**, L69 (1991)
22. Tsallis, C., Mendes, R. S. and Plastino, A. R.: The role of constraints within generalized nonextensive statistics. Physica A. **261**, 534 (1998)
23. Martınez, S., Nicolás, F., Pennini, F. et al.: Tsallis' entropy maximization procedure revisited. Physica A **286**, 489 (2000)
24. Shen, K., Zhang, B. and Wang, E.: Generalized ensemble theory with non-extensive statistics. Physica A **487**, 215 (2017)
25. Abe, S., Martınez, S., Pennini, F., et al. : Nonextensive thermodynamic relations. Phys. Lett. A **281**, 126 (2001)
26. Toral, R.: On the definition of physical temperature and pressure for nonextensive thermostatistics. Physica A **317**, 209 (2003)
27. Abe, S.: Temperature of nonextensive systems: Tsallis entropy as Clausius entropy. Physica A **368**, 430 (2006)
28. Guo, L. and Du, J. : Thermodynamic potentials and thermodynamic relations in nonextensive thermodynamics. Physica A **390**, 183 (2011)
29. Zheng, Y. and Du, J. : Nonextensive thermodynamic relations based on the assumption of temperature duality. Continuum Mech. Thermodyn. **28**, 1791 (2016)
30. Plastino, A. R. and Anteneodo, C.: A dynamical thermostatting approach to nonextensive canonical ensembles. Ann. Phys. **255**, 250 (1997)
31. Abe, S. and Rajagopal, A. K.: Nonuniqueness of canonical ensemble theory arising from microcanonical basis. Phys. Lett. A **272**, 341 (2000)
32. Garcia-Morales, V. and Pellicer, J.: Microcanonical foundation of nonextensivity and generalized thermostatistics based on the fractality of the phase space. Physica A **361**, 161 (2006)
33. Treumann, R. A. and Jaroschek, C. H.: Gibbsian theory of power-law distributions. Phys. Rev.

Lett. **100**, 155005 (2008)
34. Ruseckas, J.: Canonical ensemble in non-extensive statistical mechanics. Physica A **447**, 85 (2016)
35. Ruseckas, J.: Canonical ensemble in non-extensive statistical mechanics, q> 1. Physica A **458**, 210 (2016)
36. Abe, S.: Temperature of nonextensive systems: Tsallis entropy as Clausius entropy. Physica A **368**, 430 (2006)
37. Zheng, Y., Du, J. and Liang, F.: Thermodynamic stability criterion and fluctuation theory in nonextensive thermodynamics. Continuum Mech. Thermodyn. **30**, 629 (2018)
38. Silva, R. and Alcaniz, J. S.: Negative heat capacity and non-extensive kinetic theory. Phys. Lett. A **313**, 393 (2003)
39. Lima, J. A. S., Silva, R. and Plastino, A. R.: Nonextensive thermostatistics and the H theorem. Phys. Rev. Lett. **86**, 2938 (2001)
40. Zheng, Y. and Du, J.: The stationary state and gravitational temperature in a pure self-gravitating system. Physica A **420**, 41 (2015)
41. Zheng, Y. and Du, J.: The equivalence of isothermal and non-isothermal power law distributions with temperature duality. Physica A **427**, 113 (2015)
42. Zheng, Y.: The nonextensive parameter as a stability criterion of convection in a fluid. EPL **101**, 29002 (2013)
43. Botet, R., Płoszajczak, M., Gudima, K. K., and et al.: The thermodynamic limit in the non-extensive thermostatistics. Physica A **344**, 403 (2004)
44. Zheng, Y. and Du, J.: An application of nonextensive parameter: the nonextensive gas and real gas. Inter. J. Mod. Phys. B **21**, 947 (2007)
45. Lenzi, E. K., Mendes, R. S. and Rajagopal, A. K.: Quantum statistical mechanics for nonextensive systems. Phys. Rev. E **59**, 1398 (1999)
46. Zheng, Y.: An insight to the nonextensive parameter in the actual gas. Physica A **392**, 2487 (2013)
47. Büyükiliç, F., Demirhan, D. and Güleç, A.: A statistical mechanical approach to generalized statistics of quantum and classical gases. Phys. Lett. A **197**, 209 (1995)
48. Wang, Q. A., Nivanen, L. and Le Méhauté, A.: Generalized blackbody distribution within the dilute gas approximation. Physica A **260**, 490 (1998)